\documentclass[11pt,a4paper,notitlepage]{article}
\usepackage[utf8]{inputenc}
\usepackage[english]{babel}
\usepackage[T1]{fontenc}
\usepackage{hyperref}
\usepackage{amsmath}
\usepackage{amsfonts}
\usepackage{color} 
\usepackage{caption}
\newcommand{\Tr}{\mathrm{Tr}}
\newcommand{\tr}{\mathrm{tr}}
\newtheorem{theorem}{Theorem}

\newtheorem{lemma}{Lemma}
\usepackage{multicol}
\usepackage{amssymb}
\usepackage{graphicx}
\usepackage[left=2.5cm,right=2.5cm,top=2.5cm,bottom=2.5cm]{geometry}
\begin{document}
\begin{center}
\textbf{\Large{Constructive Tensorial Group Field Theory \\ I:
The $U(1)$-$T^4_3$ Model}}
\end{center}

\begin{center}
\vspace{20pt}

Vincent Lahoche
\footnote{vincent.lahoche@th.u-psud.fr}\\
\vspace{5pt}
{\it Laboratoire de Physique Th\'eorique, CNRS-UMR 8627, Universit\'e Paris-Sud 11, 91405 Orsay Cedex, France}
\end{center}
\vspace{10pt}

\begin{abstract}
The Loop Vertex Expansion (LVE) is a constructive technique using canonical combinatorial tools. It works well for quantum field theories without renormalization, which is the case of the field theory studied in this paper. Tensorial Group Field Theories (TGFT) are a new class of field theories 
proposed to quantize gravity. This paper is devoted to a very simple TGFT for rank three tensors 
with U(1) group and quartic interactions, hence nicknamed  $U(1)$-$T^4_3$. It has no ultraviolet divergence, and we show, with the LVE, that it is Borel summable in its coupling constant. 
\end{abstract}
\tableofcontents
\pagebreak

\section{Introduction}

Constructive field theory resums perturbative quantum field theory in order to obtain a rigorous definition of quantities such as Schwinger functions for interacting models \cite{rivbook}. The Loop Vertex Expansion (LVE) is a constructive technique \cite{constructive,constructivetensor1,constructivetensor2}, improving on the traditional constructive tools in order to treat more general models 
with non-local interactions and/or on more general geometries. Following \cite{resum}, it
 can be described as a reorganization of the perturbative series, combining an intermediate field decomposition with replicas and a forest formula. 
It allows to write the connected Schwinger functions as convergent sums indexed by spanning trees rather than as divergent sums indexed by Feynman graphs. Indeed a connected Schwinger function $S$ is usually expanded in term of a Feynman series as
\begin{equation}
S=\sum_{\mathcal{G}}\mathcal{A}_{\mathcal{G}} ,
\end{equation}
where $\mathcal{A}_{\mathcal{G}}$ is the Feynman amplitude associated to the graph $\mathcal{G}$. However, even if each of these amplitudes are ultra-violet convergent, the sum is generally badly divergent, because of the very large number of graphs of large size, so that the perturbative expansion has a zero radius of convergence in the coupling(s), hence 
\begin{equation}
\sum_{\mathcal{G}}|\mathcal{A}_{\mathcal{G}}|=\infty  .
\end{equation}
The LVE allows to circumvent this difficulty. The first step is to consider
for any pair made of a connected graph $\mathcal{G}$ and of a spanning tree  $\mathcal{T} \subset \mathcal{G}$ in it, a universal \textit{non trivial weight} $\mathit{w}(\mathcal{G},\mathcal{T})$, which is the percentage of Hepp's sectors of $G$ in which $T$ is leading, in the sense of Kruskal \emph{greedy algorithm} (see \cite{resum} for details). These weights, being by definition \emph{percentages}, are normalized:
\begin{equation}
\sum_{\mathcal{T}\subset\mathcal{G}}\mathit{w}(\mathcal{G},\mathcal{T})=1.
\end{equation}
They allow to rewrite the Feynman expansion as a sum indexed by spanning trees:
\begin{equation}
S=\sum_{\mathcal{G}}\mathcal{A}_{\mathcal{G}}=\sum_{\mathcal{G}}\sum_{\mathcal{T}\subset\mathcal{G}}\mathit{w}(\mathcal{G},\mathcal{T})\mathcal{A}_{\mathcal{G}}=\sum_{\mathcal{T}}\mathcal{A}_{\mathcal{T}},
\end{equation}
where:
\begin{equation}
\mathcal{A}_{\mathcal{T}}:=\sum_{\mathcal{G}\supset\mathcal{T}}\mathit{w}(\mathcal{G},\mathcal{T})\mathcal{A}_{\mathcal{G}}. \label{graphsum}
\end{equation}
Since trees do not proliferate as fast as Feynman graphs, in good cases it can be shown that:
\begin{equation}
\sum_{\mathcal{T}}|\mathcal{A}_{\mathcal{T}}|<\infty  \label{treesum}
\end{equation}
at least in a certain domain that we call the \textit{summability domain}. Strictly speaking, such a program can be achieved 
with standard Feynman graphs only for Fermionic theories, because of the Pauli principle, since in that case amplitudes 
at same order combine into a determinant implying nice compensations \cite{ARfermio}. Such compensations do not occur  at fixed order 
for Bosonic theories, hence the sum \eqref{treesum} does not converge, even if it is repacked as a tree expansion. Fortunately the loop vertex expansion
overcomes this difficulty by working in another representation, called intermediate field, or Hubbard-Stratonovic. This representation
amounts to a clever exchange of the roles of propagators and vertices. 
The program summarized by equations \eqref{graphsum}-\eqref{treesum}    then works, but \emph{for
the graphs of the intermediate field representation}, and
the corresponding sum \eqref{treesum} converges absolutely to the \emph{Borel sum} of the initial expansion. \\

Group Field Theories (GFT), on the other hand, are a class of field theories defined on a group manifold and characterized by a specific form of non-locality in their interactions, giving their Feynman diagrams the structure of cellular complexes rather than graphs \cite{GFTreviews}. In some recent works, such theories appear as a promising way for research on quantum gravity. More precisely, GFTs can be viewed either as a tentative to extend to higher dimensions the success of matrix models in dimension two \cite{matrix}, or as a second quantization of loop quantum gravity states
because spin foams arise as Feynman diagrams of GFTs \cite{GFT-LQG}.
Tensorial Group Field Theories (TGFTs) are a new class of GFTs, whose propagator is based on an inverse Laplacian \cite{Laplacian}
and for which interactions are chosen to be \textit{invariant}
\cite{Rivasseau-track,TGFTrenorm-Joseph,TGFTrenorm-Carrozza,TGFTrenorm-others,Lahoche:2015ola}, in the precise sense of the $U(N)^{\otimes d}$ invariance 
of rank-$d$ tensors of  size $N$ \cite{expansion,Bonzom:2011zz,uncoloring,var-tens}. This $U(N)^{\otimes d}$ invariance provides a $1/N$ expansion. It requires tensor indices to be contracted into pairs \emph{respecting position of indices}\footnote{This rule is the main imprvement over previous more singular tensor models \cite{tensor}.}.
The same scheme is used for TGFTs, the sums overs indices being replaced by integrations over group variables. For TGFTs, it has been proved recently that this additional invariance allows to define a locality principe, the \textit{traciality}, useful for renormalization and for 
importing other classical field theory tools, such as the functional renormalization group \cite{EichhornKoslowski,BBGO,Geloun:2015qfa}
Renormalization and phase transitions are at the core of the space-time emergence and \textit{geometrogenesis scenario} \cite{PTO}. Space-time emergence should correspond to a phase transition from a symmetric to a \textit{condensed} phase, similar to the Bose-Einstein condensation in physics of many-bodies systems. As recent works seems to confirm, \textit{asymptotic freedom} \cite{TGFTrenorm-Joseph,TGFTrenorm-others,Dine,Rivasseau-AF} phase transitions \cite{Delepouve:2015nia,Benedetti:2015ara}
 and \textit{infrared non-Gaussian fixed points} \cite{Lahoche:2015ola,BBGO,Geloun:2015qfa,Carrozza:2014rya} are common features of TGFTs, and promising steps in the long way towards understanding semi-classical space-time.\\

In this paper we propose a systematic constructive study of TGFTs, starting with the simplest non-trivial model, namely tensors of rank
3, with  Abelian group U(1), and the simplest color-symmetric quartic interactions. We call the corresponding model the $U(1) - T^4_3$ model.
With the LVE technique, we prove Borel summability in the coupling constant of its free energy and connected Schwinger functions. As we
shall see, the \textit{closure constraint} coming from the GFTs origins of our model, plays an important role, because it reduces the effective intermediate fields, which are matrices, to their diagonal parts.

For more complicated TGFT models which require renormalization, 
the LVE must be generalized to a multiscale expansion called the MLVE \cite{MLVE}. The first model of this kind, the rank-four $U(1) - T^4_4$
model will be the subject of a companion paper.

\section{The model}
\subsection{A rank 3 ultra-violet convergent Abelian TGFT}

A Group Field Theory (GFT) is a non local field theory defined on $\mathrm{G}^{d}$, where $\mathrm{G}$ is a Lie group. A field $\psi$ is therefore a map $\psi:\mathrm{G}^d\to\mathbb{C}$, but, very importantly, it can also be considered as a rank-$d$ tensor whose indices take values in the the group $\mathrm{G}$. The field theory is defined by a generating functional integral:
\begin{equation}\label{partitionfunc}
\mathcal{Z}[J,\bar{J}]=\int d\mu_{C}(\psi,\bar{\psi})e^{-S_{int}[\psi,\bar{\psi}]+\langle \bar{J},\psi\rangle+\langle \bar{\psi},J\rangle}.
\end{equation}
In this equation, $J$ and $\bar{J}$ are understood as external sources, respectively associated to the fields $\psi$ and $\bar{\psi}$. As these fields, the sources are maps from $\mathrm{G}^d$ to $\mathbb{C}$, and the notation $\langle \bar{J},\psi\rangle$ means that:
\begin{equation}
\langle \bar{J},\psi\rangle:=\int [dg]^d \bar{J}(g_1,...,g_d)\psi(g_1,...,g_d),
\end{equation}
where $dg$ is the Haar measure over the group. In addition to the source terms, there are two pieces in the definition \ref{partitionfunc}. The first one is the Gaussian measure $d\mu_C$, requiring the choice of a covariance $C$. The second piece is the interaction called $S_{int}$. The theory is said to be ``tensorial" if these terms are invariant under independent unitary transformations on each tensor index. The only way to achieve this is that, in each terms of $S_{int}$, any variable attached to a field $\psi$ is identified to another variable of a field $\bar{\psi}$ \emph{with same position}, and summed over the group manifold. An example in $d=3$ is:
\begin{align}\label{intexample}
\int \prod_{i=1}^3 dg_idg_i'&\psi(g_1,g_2,g_3)\bar{\psi}(g_1',g_2,g_3)\psi(g_1,g_2',g_3')\bar{\psi}(g_1',g_2',g_3').
\end{align}
Such a interaction, with tensorial invariance, is called \textit{bubble} and a GFT with interactions of this type is called a TGFT. Each bubble can be pictured graphically as a $3$-colored bipartite regular graph, where Fields $\psi$ and $\bar{\psi}$ are represented respectively by white and black vertices. Each of these vertices are joined by $d$ lines, representing the group variables of the field, and the way according to which these lines are connected to the vertices follows the sums over field variables. An example is given by figure \ref{fig1} for the bubble \ref{intexample}. \\
\begin{center}
\includegraphics[scale=1]{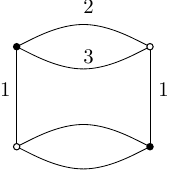} 
\captionof{figure}{Bipartite graph associated to the interaction \ref{intexample}.}\label{fig1}
\end{center}

In this paper we\label{sectionmodel} shall consider only a specific model of TGFT on the compact Abelian group $U(1)^3$. Hence a field $\psi$ is a map $\psi:U(1)^3\to\mathbb{C}$. Furthermore in our model, fields are chosen to be invariant under right uniform group translation of each variable:
\begin{equation}\label{gaugesym}
\psi(g_1,g_2,g_3)=\psi(g_1h,g_2h,g_3h)\qquad \forall h\in U(1). 
\end{equation}
This symmetry, called ``closure constraint" can be understood as a discrete gauge invariance in the language of lattice gauge theory. It comes from simplicial gravity and spin foam model, where it is interpreted as a discrete connection, and can be seen as an additional enrichment or pre-geometric data for TGFTs \cite{GFTreviews,GFT-LQG}. At the field theory level, this symmetry is implemented by the Gaussian measure, and we choose the following propagator:
\begin{align}
\int &d\mu_{C}(\psi,\bar{\psi})\bar{\psi}(g_1,g_2,g_3)\psi(g'_1,g'_2,g'_3)=\int_{U(1)}dh\int_{0}^{\infty}d\alpha e^{-\alpha m^2}\prod_{i=1}^3K_{\alpha}(g_ihg_i^{\prime\,-1}),
\end{align}
where $K_{\alpha}$ is the heat kernel over $U(1)$, satisfying the heat kernel equation:
\begin{equation}
\Big(\frac{\partial}{\partial \alpha}-\Delta\Big)K_{\alpha}=0 .
\end{equation}
We restrict to quartic interactions (which are all 
automatically \emph{melonic} \cite{Gurau:2011xp} in rank 3). An example is given by \ref{intexample}, and pictured on \ref{fig1}. More precisely, we choose:
\begin{equation}\label{int}
S_{int}[\bar{\psi},\psi]=\lambda\sum_{i=1}^3\Tr_{b_i}[\bar{\psi},\psi]
\end{equation}
where the coupling constant $\lambda$ is a complex number, and $\Tr_{b_i}$, the ``invariant trace" over the bubble $b_i$ means the contraction of field variables involved in the trace, following the scheme of the bubble $b_i$, the index $i$ referring to the color of the intermediate line (hence, figure \ref{fig1} and formula \ref{intexample} correspond to $b_1$). \\

Schwinger  $N$-points functions can be defined by their Feynman expansion in power of $\lambda$:
\begin{equation}
G_N=\sum_{\{\mathcal{G}_N\}}\frac{(-\lambda)^{V(\mathcal{G}_N)}}{s(\mathcal{G}_N)}\mathcal{A}_{\mathcal{G}_N}
\end{equation}
where $\{\mathcal{G}_N\}$ is the set of graphs with $N$ external lines, $V(\mathcal{G})$ the number of vertices in $\mathcal{G}$, $s(\mathcal{G})$ a symmetry factor, and $\mathcal{A}_{\mathcal{G}_N}$ the amplitude, whose explicit expression can be obtained from Feynman rules:
\begin{align}
\nonumber\mathcal{A}_{\mathcal{G}}= \int_{U(1)^{L(\mathcal{G})}} \int_{0}^{\infty}\prod_{e\in \mathcal{L}(\mathcal{G})}{d\alpha_{e} e^{-\alpha_{e} m^{2}}} {dh_{e}}\label{amplitude}\prod_{f\in \mathcal{F}_{ext}(\mathcal{G})}&K_{\alpha_{(f)}} {\left( g_{s(f)}\vec{\prod}_{e\in\partial{f}}h_{e}^{\epsilon_{ef}}g^{-1}_{t(f)} \right)}\\
&\times{ \prod_{f\in \mathcal{F}(\mathcal{G})} K_{\alpha_{(f)}} {\left( \vec{\prod}_{e\in\partial{f}}h_{e}^{\epsilon_{ef}} \right)}}.
\end{align}
where $\mathcal{L}(\mathcal{G})$, $\mathcal{F}(\mathcal{G})$ and $\mathcal{F}_{ext}(\mathcal{G})$ are respectively the sets of lines, internal and external faces, and $L(\mathcal{G})$ is the cardinal of $\mathcal{L}(\mathcal{G})$. Finally $\epsilon_{fe}$ is the incidence matrix, whose elements are equal to $\pm 1$ or $0$ if the line $e$ is in the boundary of $f$, noted $\partial f$, or not (the sign depends on the relative orientation of the line with respect to the orientation of the face)\footnote{Interestingly, the incidence matrix can be understood as the boundary operator for the $2$-complex build as the set of vertices, lines and faces of $\mathcal{G}$.}. Note that in the language of lattice gauge theory, the integrals over $h_e$ group variables in the previous amplitude \ref{amplitude} can be understood as a sum over discrete connections on a cellular complex defined by the Feynman graph $\mathcal{G}$. In this point of view, the product $\prod_{e\in\partial f}h_e^{\epsilon_{ef}}$ are holonomies around the face $f$, and the $G^{\times |V(\mathcal{G})|}$ symmetry of the Feynman amplitudes (where $|V(\mathcal{G})|$ is the vertex number of $\mathcal{G}$), coming from  invariance under the  transformations
\begin{equation}
h_e\to k_{s(e)}h_ek^{-1}_{t(e)} \qquad k_v\in G,
\end{equation}
is a discrete version of gauge invariance.
\begin{center}
\includegraphics[scale=1]{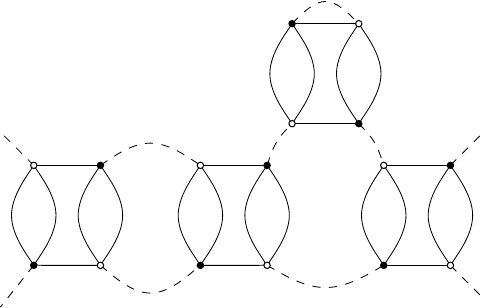}
\captionof{figure}{Example of Feynman graph with $4$ external lines.}\label{fig2} 
\end{center}
A Feynman graph can be pictured graphically with bubbles (interaction vertices) pictured as in figure \ref{fig1}, and Wick contractions  pictured as dotted lines joining two black and white vertices, either belonging to the same bubble or not. Figure \ref{fig2} gives an example of such a Feynman graph. As a result, a Feynman graph is also a bipartite colored graph, with black and white vertices and colored lines, whose color runs from $0$ to $3$, the color $0$ being associated to the dotted lines.\\

\noindent
The heat kernel over $U(1)$ can be easily found using Fourier decomposition, as:
\begin{equation}
K_{\alpha}(\theta)=\sum_{p\in\mathbb{Z}}e^{ip\theta}e^{-\alpha p^2},
\end{equation}
where $\theta$ is the coordinate of the element $e^{i\theta}\in U(1)$. It follows that in momentum representation, the propagator is written as:
\begin{equation}
C_{\vec{p},\vec{p}^{\,\prime}}=\delta_{\vec{p},\vec{p}^{\,\prime}}\dfrac{\delta\big(\sum_ip_i\big)}{\vec{p}^2+m^2}
\end{equation}
where $\vec{p}:=(p_1,p_2,p_3)\in \mathbb{Z}^3$, is a convenient notation for the set of Fourier variables. 

\subsection{Power counting and melons}

Power counting has been established for such field theories in recent works \cite{TGFTrenorm-Carrozza,TGFTrenorm-others}, using standard multi-scale analysis, and it has been proved that the divergent degree $\omega(\mathcal{G})$ for a Feynman graph $\mathcal{G}$ is given by:
\begin{equation}
\omega(\mathcal{G})=-2L(\mathcal{G})+F(\mathcal{G})-R(\mathcal{G}).
\end{equation}
In this formula, $L(\mathcal{G})$ and $F(\mathcal{G})$ are respectively the number of lines and faces of the Feynman graph $\mathcal{G}$, where a ``face" is defined as a cycle of lines, and $R(\mathcal{G})$ is the rank of the incidence matrix $\epsilon_{fe}$.\\

As explained before, the leading order contributions come from the set of melonic graphs \cite{expansion1,expansion2,expansion3}, which can be defined recursively in a purely graphical framework as follows. The elementary melon with four colors and minimum number of black and white vertices - that is to say $2$ - is the so-called ``supermelon" graph, pictured on figure \ref{fig3} below, where the dotted line is associated to the color $0$.
\begin{center}
\includegraphics[scale=1.2]{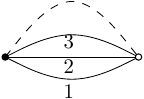} 
\captionof{figure}{The supermelon graph.}\label{fig3}
\end{center}
Then, from a melon with $2p$ black and white vertices, we build a melon with $2(p+1)$ vertices by replacing a line of color $i=0,...,3$ of the starting graph of order $2p$ with the so-called elementary melon of color $i$ pictured in figure \ref{fig4} for color $1$. 
\begin{center}
\includegraphics[scale=1.2]{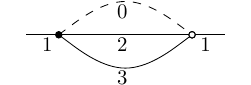}
\captionof{figure}{The elementary melon of color $1$, replacing a line of color $1$ in the recursion.}\label{fig4}
\end{center}
In summary, a generic melonic graph has a matriochka structure, consisting in elementary melons nested into elementary melons, and so on, and an example is given in figure \ref{fig5}. As a result, we see that our interaction in figure \ref{fig1} has exactly the structure of a melonic graph with three colors, justifying the terminology of melonic interaction. 
\begin{center}
\includegraphics[scale=0.8]{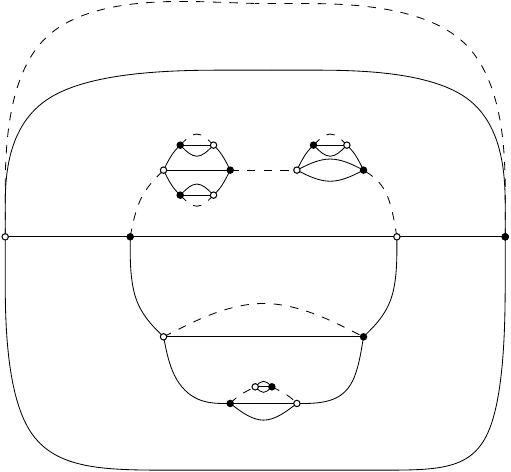} 
\captionof{figure}{An example of melonic graph}\label{fig5}
\end{center}
An example of melonic Feynman graph in the field theory framework of \ref{sectionmodel} is given in section \ref{fig6}. 
\begin{center}
\includegraphics[scale=1.2]{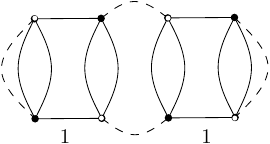} 
\captionof{figure}{Example of melonic Feynman vacuum graph with two bubble vertices.}\label{fig6}
\end{center}
Returning to our field theory, it is easy to see that for any non-vacuum melonic Feynman graph \cite{TGFTrenorm-Carrozza}:
\begin{equation}
F(\mathcal{G})-R(\mathcal{G})=L(\mathcal{G})-V(\mathcal{G})+1,
\end{equation}
so that, taking into account of the relationship $L(\mathcal{G})=2V(\mathcal{G})-N(\mathcal{G})/2$, where $N$ is the number of external lines, we find that the divergent degree write as:
\begin{equation}
\omega(\mathcal{G})=1+\dfrac{N(\mathcal{G})}{2}-3V(\mathcal{G}).
\end{equation}
Hence, because the number of external lines of a graph with $V(\mathcal{G})$ vertices is bounded by $N_{max}=2(V(\mathcal{G})+1)$, when the graph is a tree, without faces, the divergent degree is strictly smaller than $\omega_{max}=2(1-V(\mathcal{G}))\leq 0$, which is strictly smaller than $0$ for $V>1$. For $V=1$ our bound could indicate a logarithmic divergence, but a direct calculation shows that the divergent degree of a melonic graph with one vertex, as pictured below, is $$\omega_{melo}(V=1)=-2+2-1=-1,$$ so that our model is in fact divergence free, at least for non-vacuum graphs. 
For vacuum graphs, the following bound can be established \cite{TGFTrenorm-Carrozza,TGFTrenorm-others}  (saturated for melonic graphs):
\begin{equation}
F(\mathcal{G}_{vac})-R(\mathcal{G}_{vac})\leq L(\mathcal{G}_{vac})-V(\mathcal{G}_{vac})+2
\end{equation}
so that, 
\begin{equation}
\omega_{melo}(\mathcal{G}_{vac})=2-3V(\mathcal{G}_{vac})<0.
\end{equation}
Hence this model is totally free of ultraviolet divergences.

\subsection{Intermediate field formalism}

\label{section23}
The intermediate field formalism is the first ingredient of the LVE \cite{constructive,constructivetensor,TGFTrenorm-others}. In Fourier component, the action \ref{int} can be written as:
\begin{equation}\label{int2}
S_{int}=\lambda\sum_{i=1}^3\sum_{\{\vec{p}_i\}}\mathcal{W}^{(i)}_{\vec{p_1},\vec{p_2},\vec{p}_3,\vec{p}_4}T_{\vec{p}_1}\bar{T}_{\vec{p}_2}T_{\vec{p}_3}\bar{T}_{\vec{p}_4},
\end{equation}
where $T_{\vec{p}}$ (resp $\bar{T}_{\vec{p}}$) are the Fourier components of the field $\psi$ (resp $\bar{\psi}$): $\psi(\vec{\theta})=\sum_{\vec{p}\in\mathbb{Z}}T_{\vec{p}}e^{i\vec{p}\cdot\vec{\theta}}$, and the symbols $\mathcal{W}^{(i)}$ are products of Kronecker deltas:
\begin{equation}
\mathcal{W}^{(i)}_{\vec{p_1},\vec{p_2},\vec{p}_3,\vec{p}_4}=\delta_{p_{1i}p_{4i}}\delta_{p_{2i}p_{3i}}\prod_{j\neq i}\delta_{p_{1j}p_{2j}}\delta_{p_{3j}p_{4j}}.
\end{equation}
Hence, defining the three Hermitian matrices $\mathbb{M}^i$ with elements
\begin{equation}
\mathbb{M}^i_{mn}:=\sum_{\{\vec{p}_1,\vec{p}_2\}}\prod_{j\neq i}\delta_{p_{1j}p_{2j}}\delta_{p_{1i}n}\delta_{p_{2i}m}T_{\vec{p}_1}\bar{T}_{\vec{p}_2},
\end{equation}
the action \ref{int2} can be rewritten in term of these matrices $\mathbb{M}^{i}$:
\begin{equation}
S_{int}=\lambda\sum_{i=1}^3\tr(\mathbb{M}^{i})^2
\end{equation}
where ``$\tr$" mean the trace over indices of the matrices $\mathbb{M}^{i}$. The intermediate field decomposition arises as an application of the well known properties of the Gaussian integration to the partition function \ref{partitionfunc}:
\begin{align}\label{intermediate1}
\mathcal{Z}[J,\bar{J}]&=\int d\mu_{C}(\psi,\bar{\psi})e^{-\lambda\sum_{i=1}^3\tr(\mathbb{M}^{i})^2+\langle \bar{J},\psi\rangle+\langle \bar{\psi},J\rangle}\\
&=\frac{\int \prod_{i=1}^3d\sigma_ie^{-\tr(\sigma_i)^2}}{\int \prod_{i=1}^3d\sigma_i'e^{-\tr(\sigma_i')^2}}\int d\mu_{C}(\psi,\bar{\psi})e^{i\sqrt{2\lambda}\sum_{i=1}^3\tr(\sigma_i\mathbb{M}^{i})+\langle \bar{J},\psi\rangle+\langle \bar{\psi},J\rangle}\\
&=\int d\nu_{\mathbb{I}}(\sigma)e^{-\Tr\ln(1-i\sqrt{2\lambda}C\Sigma)-\bar{J}RJ}
\end{align}
where in the last line the integration over $\bar{\psi},\psi$ has been performed. $R:=(1-i\sqrt{2\lambda}C\Sigma)^{-1}C$ is the \textit{resolvent matrix}, $d\nu_{\mathbb{I}}(\sigma)$ is the normalized Gaussian integration over the $\sigma_i$ with covariance the identity matrix 
$\mathbb{I}$ (Gaussian unitary ensemble), and:
\begin{equation}
\Sigma:=     \sigma_1  \otimes   \mathbb{I}   \otimes   \mathbb{I}   +    \mathbb{I}   \otimes  \sigma_2\otimes  \mathbb{I}  
+  \mathbb{I}   \otimes   \mathbb{I} \otimes  \sigma_3 .
\end{equation}
Perturbatively, we can expand the big trace of the logarithm and the resolvent as
\begin{equation}
\Tr\ln(1-i\sqrt{2\lambda}C\Sigma)=\sum_n\frac{1}{n}\Tr(i\sqrt{2\lambda}C\Sigma)^n\,;\quad (1-i\sqrt{2\lambda}C\Sigma)^{-1}C=\sum_n(i\sqrt{2\lambda}C\Sigma)^nC.
\end{equation}
We call the interactions $\Tr(i\sqrt{2\lambda}C\Sigma)^n$ \textit{loop vertices}, and $(i\sqrt{2\lambda}C\Sigma)^nC$ \textit{ciliated vertices} \cite{TGFTrenorm-others}. Graphically, the loop vertex $\Tr(i\sqrt{2\lambda}C\Sigma)^n$ is pictured by a grey disk with $n$ half colored wavy lines, the ciliated vertex $(i\sqrt{2\lambda}C\Sigma)^nC$ by a lighter grey disk with $n$ half colored wavy lines and one dotted cilium (which represent the external field line), and the propagators of the intermediate fields are represented by colored wavy lines, joining  the half wavy lines of the grey disks, accordingly to each proper colors (see \cite{TGFTrenorm-others}). Figure \ref{fig7} gives an example of Feynman graph in intermediate field representation. Note that the dotted lines of the original representation given by \ref{partitionfunc} are in one-to-one correspondence with the \textit{arcs} of the grey disks in the intermediate field decomposition.  
\begin{center}
\includegraphics[scale=0.8]{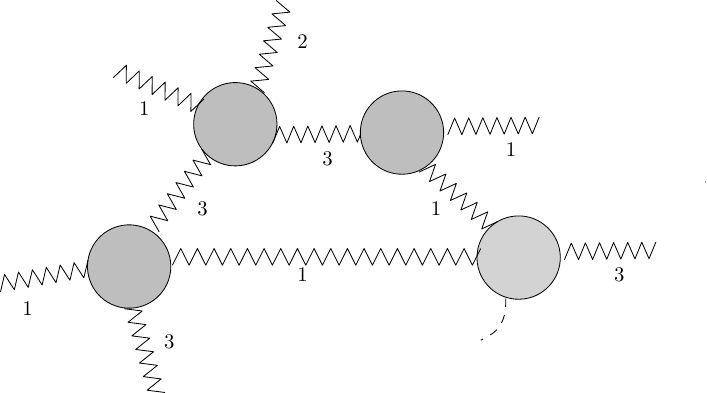}
\captionof{figure}{Example of Feynman graph in intermediate field representation, with one ciliated vertex.}\label{fig7}
\end{center}
We conclude this section by underlining an important simplification coming from gauge invariance \ref{gaugesym} already mentioned in the literature \cite{TGFTrenorm-others}. If we call $\tau_i(p_i):=(\sigma_i)_{p_ip_i}$ the \textit{diagonal part} of matrix $\sigma_i$, we have:
\begin{equation}
\Tr(i\sqrt{2\lambda}C\Sigma)^n=(i\sqrt{2\lambda})^n \sum\limits_{\substack{\vec{p}\in\mathbb{Z}^3}{\vec{k}\in\mathbb{N}^3|\sum_ik_i=n}}\frac{n!}{\prod_lk_l!}\dfrac{\delta\big(\sum_ip_i\big)}{(\vec{p}^2+m^2)^n}\prod_{i=1}^3[\tau_i(p_i)]^{k_i},
\end{equation}
and a similar result holds for the ciliated vertices \cite{TGFTrenorm-others}. Hence, only the diagonal part of the intermediate field contribute effectively, so that we can reduce our three intermediate field matrices $\sigma_i$ to three \textit{intermediate vector fields} $\tau_i$, so that \ref{intermediate1} writes:
\begin{equation}\label{intermediatevec}
\mathcal{Z}[J,\bar{J}]=\int d\nu_{\mathbb{I}}(\tau)e^{-\sum_{\vec{p}\in\mathcal{P}}\ln(1-i\sqrt{2\lambda}C_0(\vec{p})\Gamma(\vec{p}))-\sum_{\vec{p}\in\mathcal{P}}\bar{J}(\vec{p})(1-i\sqrt{2\lambda}C_0(\vec{p}))\Gamma(\vec{p}))^{-1}C_{0}(\vec{p})J(\vec{p})},
\end{equation}
where $C_0(\vec{p}):=(\vec{p}^2+m^2)^{-1}$, $\mathcal{P}:=\{\vec{p}\in\mathbb{Z}^3|\sum_ip_i=0\}$, $\Gamma(\vec{p}):=\sum_i\tau_i$, and $d\nu_{\mathbb{I}}(\tau)$ is the Gaussian measure of the three vector fields, defined as:
\begin{equation}
\int d\nu_{\mathbb{I}}(\tau) \tau_i(p)\tau_j(p'):=\delta_{ij}\delta_{pp'}.
\end{equation}

\section{BKAR Forest formula and Borel summability}
\subsection{The ``constructive swiss knife"}

The BKAR (Brydges–Kennedy–Abdesselam–Rivasseau) forest interpolation formula \cite{BKAR}, nicknamed the ``constructive swiss knife", is the heart of the LVE. A forest formula expands a quantity defined on $n$ points in terms of forests built on these points, and is a multi-variable Taylor expansion with integral remainder. There are in fact many forest formulas, but the BKAR formula seems the only one which is both \textit{symmetric} under permutation of the $n$ points and \textit{positive} \cite{constructive}. \\

Let $[1, \cdots, n]$ be the finite set of points considered above. An edge $l$ between two elements $i,j\in [1, \cdots, n]$ is a couple $(i,j)$ for $1\leq i<j\leq n$, and the set of such edges can be identified with the set of lines of $K_n$, the complete graph with $n$ vertices. Consider the vector space $S_n$ of $n\times n$ symmetric matrices, whose dimension is $n(n+1)/2$ and the \textit{compact and convex} subset $PS_n$ of \textit{positive} symmetric matrices whose diagonal coefficients are all equal to $1$, and off-diagonal elements are between $0$ and $1$. Any $X\in PS_n$ can be parametrized by $n(n-1)/2$ elements $X_l$, where $l$ run over the edges of the complete graph $K_n$ \cite{constructive}. Let us consider a smooth function $f$ defined in the interior of $PS_n$ with continuous extensions to $PS_n$ itself. The BKAR forest formula state that:
\begin{theorem}\label{BKAR} \textbf{(The BKAR forest formula)}
\begin{equation}
f(\mathbf{1})=\sum_{\mathcal{F}}\int d\mathit{w}_{\mathcal{F}}\partial_{\mathcal{F}}f[X^{\mathcal{F}}(\mathit{w}_{\mathcal{F}})]
\end{equation}
where $\mathbf{1}$ is the matrix with all entries equal to $1$, and:\\

\noindent
$\bullet$ The sum is over the forests $\mathcal{F}$ over $n$ labeled vertices, including the empty forest.\\

\noindent
$\bullet$ The integration over $d\mathit{w}_{\mathcal{F}}$ means integration from $0$ to $1$ over one parameter for each edge of the forest. Note that there are no integration for the empty forest since by convention an empty product is $1$. \\

\noindent
$\bullet$ $\partial_{\mathcal{F}}:=\prod_{l\in\mathcal{F}}\partial_l$ means a product of partial derivatives with respect to the variables $X_l$ associated to the edge $l$ of $\mathcal{F}$. \\

\noindent
$\bullet$ The matrix $X^{\mathcal{F}}(\mathit{w}_{\mathcal{F}})\in PS_n$ is such that $X^{\mathcal{F}}_{ii}(\mathit{w}_{\mathcal{F}})=1\,\forall i$, and for $i\neq j$ $X^{\mathcal{F}}_{ij}(\mathit{w}_{\mathcal{F}})$ is the infimum of the $\mathit{w}_l$ variables for $l$ in the unique path from $i$ to $j$ in $\mathcal{F}$. If no such path exists, by definition $X^{\mathcal{F}}_{ij}(\mathit{w}_{\mathcal{F}})=0$. 
\end{theorem}

\subsection{Borel summability}

The third key ingredient  is \textit{Borel summability}, and a helpful theorem is \cite{sokal}:
\begin{theorem}\textbf{(Nevanlinna)}
A series $\sum_{n}\frac{a_n}{n!}\lambda^n$ is \textit{Borel summable} to a function $f(\lambda)$ if the following conditions are met:\\

\noindent
$\bullet$ $f(\lambda)$ is analytic in a disk $Re(\lambda^{-1})>R^{-1}$ with $R\in\mathbb{R}^+$. \\

\noindent
$\bullet$ $f(\lambda)$ admits a Taylor expansion at the origin:
\begin{equation}
f(\lambda)=\sum_{k=0}^{r-1}f_{k}\lambda^k+R_rf(\lambda), \qquad |R_rf(\lambda)|\leq K\sigma^rr!|\lambda|^r, \label{borelbound}
\end{equation}
for some constants $K$ and $\sigma$ independent of $N$.\\

\noindent
If $f(\lambda)$ is Borel summable in $\lambda$, then:
\begin{equation}
\mathcal{B}(t)=\sum_{n=0}^{\infty}\frac{1}{n!}f_{n}t^n
\end{equation}
is an analytic function for $|t|<\sigma^{-1}$ which admits an analytic continuation in the strip $\{z|\,|Im(z)|<\sigma^{-1}\}$ such that $|\mathcal{B}(t)|\leq Be^{t/R}$ for some constant $B$ and $f(\lambda)$ is represented by the absolutely convergent integral:
\begin{equation}
f(\lambda)=\frac{1}{\lambda}\int_{0}^{+\infty}dt\mathcal{B}(t)e^{-t/\lambda}.
\end{equation}
\end{theorem}\label{Borel}
The aim of the rest of this paper is to combine the intermediate field representation with the forest formula in order to obtain a tree expansion for the free energy and Schwinger functions of our theory. This expansion will be shown to converge in a cardioid domain \cite{constructivetensor}, larger than the one 
of the Nevanlinna theorem. Bound \eqref{borelbound} will be proven, hence Borel summability follows. 

\section{Convergence and summability}

This section is devoted to the main convergence and analyticity theorem using the LVE. We begin with the free energy, then extend to connected Schwinger functions. For the convenience of the reader, two lemmas with their respective proofs are reported at the end of the first subsection. 

\subsection{Free energy}

We start\label{free} with the partition function \ref{intermediatevec} and we expand to infinity the exponential of the interaction (note that sources are discarded):
\begin{equation}\label{partitionexp}
Z(\lambda)=\int d\nu_{\mathbb{I}}(\tau)\sum_{n=0}^{\infty} \frac{1}{n!}(-W(\tau))^n,
\end{equation}
where: $W:=\sum_{\vec{p}\in\mathcal{P}}\ln(1-i\sqrt{2\lambda}C_0(\vec{p})\Gamma(\vec{p}))$. The first step is to introduce a \textit{replica trick} for the Bosonic intermediate fields. We duplicate the intermediate field into copies, so that:
\begin{equation}
(-W(\tau))^n\to\prod_{m=1}^n(-W_m(\tau_m))
\end{equation}
and in the same time replace the covariance $\mathbb{I}$ by $\mathbf{1}_n$, the $n\times n$ matrix with all entries equals to $1$, so that our measure write as: $d\nu_{\mathbf{1}}(\tau_m)$. Exchanging sum and Gaussian integration, \ref{partitionexp} becomes
\begin{equation}
Z(\lambda)=\sum_{n=0}^{\infty}\frac{1}{n!}\int d\nu_{\mathbf{1}_n}(\tau_m)\prod_{m=1}^n(-W_m(\tau_m)).
\end{equation}
In order to apply the forest formula, we introduce the \textit{coupling parameters} $x_{mp}$, so that $x_{mp}=x_{pm},\, x_{pp}=1$ between the  replicas. Hence
\begin{equation}
Z(\lambda)=\sum_{n=0}^{\infty}\frac{1}{n!}\int \prod_{i,m}d\tau_{i,m}e^{-\frac{1}{2}\sum_{m,p=1}^nx_{mp}\tau_{i,m}\tau_{i,p}}\prod_{m=1}^n(-W_m(\tau_m))\bigg|_{x_{pm}=1}
\end{equation}
where the first indice of $\tau_{i,m}$ is the color index (running from $1$ to $3$), and the second is the replica index (running from $1$ to $n$). Note that in our notation, the sum over momenta in the expression of the Gaussian measure is implied. Applying the BKAR forest formula, we find, in derivative representation of Gaussian integration \cite{constructive}:
\begin{align}\label{expZ}
Z(\lambda)=\sum_{n=0}^{\infty}\frac{1}{n!}\sum_{\mathcal{F}_n}\int_{0}^1\bigg(\prod_{l\in\mathcal{F}_n}d\mathit{w}_l\bigg)\Bigg[&e^{\frac{1}{2}\sum_{m,p=1}^nX_{mp}(\mathit{w}_l)\frac{\partial}{\tau_{i,m}}\frac{\partial}{\tau_{i,p}}}\\\nonumber
&\qquad \times\prod_{l\in \mathcal{F}} \bigg(\dfrac{\partial^2}{\partial \tau_{is(l)}\partial \tau_{it(l)}}\bigg)\prod_{m=1}^n(-W_m(\tau_m))\Bigg]_{\tau_{im}=0}
\end{align}
where $s(t)$ and $t(l)$ are respectively the replica indices for source and target of the edge $l$. As well known in quantum and statistical field theory, the \textit{free energy} $F:=\ln(Z)$ expands as a sum over amplitudes labeled by connected Feynman graphs. Because the expansion 
\eqref{expZ} factorizes over the connected components of the forest and since the connected version of a forest is a tree, we obtain 
the following tree expansion for the free energy $F$:
\begin{align}\label{treeexp1}
F-F_0=\sum_{\mathcal{T}}\frac{1}{V(\mathcal{T})!}\int_{0}^1\bigg(\prod_{l\in\mathcal{T}}d\mathit{w}_l\bigg)&\Bigg[\int d\nu_{X(\mathit{w}_l)}(\tau)\prod_{l\in \mathcal{T}} \prod_{i=1}^3\bigg(\dfrac{\partial^2}{\partial \tau_{is(l)}\partial \tau_{it(l)}}\bigg)\prod_{m=1}^{V(\mathcal{T})}(-W_m(\tau_m))\Bigg]
\end{align}
where $V(\mathcal{T})$ is the vertex number of the tree $\mathcal{T}$, and $F_0$ is defined as the contribution to the sum \ref{expZ} with $n=1$:
\begin{equation}
F_0:=\int d\nu_{X(\mathit{w}_l)}(\tau)\sum_{\vec{p}\in\mathcal{P}}\ln(1-i\sqrt{2\lambda}C_0(\vec{p})\Gamma(\vec{p})).
\end{equation}
Note that the replica indices runs now from $1$ to $V(\mathcal{T})$. The derivatives can be performed, to get:
\begin{align}\label{sumtree}
F-F_0=\sum_{\mathcal{T}}&\frac{(-2\lambda)^{V(\mathcal{T})-1}}{V(\mathcal{T})!}\int_{0}^1\bigg(\prod_{l\in\mathcal{T}}d\mathit{w}_l\bigg)\Bigg[\int d\nu_{X(\mathit{w}_l)}(\tau)\prod_{m=1}^{V(\mathcal{T})}\\
&\qquad \times\sum_{\vec{p}_m\in\mathcal{P}}(c(m)-1)!R(\vec{p}_m)^{c(m)}\prod_{l\in\mathcal{T}}\bigg(\sum_{i=1}^3\delta_{p_{t(l)i}p_{s(l)i}}\bigg)\Bigg]
\end{align}
where $c(m)$ is the set of \textit{arcs} of the vertex $m$ and $R(\vec{p})$ is the resolvent of section \ref{section23} (in momentum representation):
\begin{equation}
R(\vec{p})=\frac{C_0(\vec{p})}{1-i\sqrt{2\lambda}C_0(\vec{p})\Gamma(\vec{p})}.
\end{equation}
Each contribution to the sum over trees \ref{sumtree} can be pictured as in figure \ref{fig8} below, where each \textit{disk} or vertex represents a product over arc-resolvents with the same variable $\vec{p}_m$, and the labeled $1,2,3$ lines represent the identification of internal index between two (necessarily different) vertices, coming from the Kronecker deltas $\delta_{p_{t(l)i}p_{s(l)i}}$. The labels of the lines refer to the color $i$ of these Kronecker delta, and the number of lines starting from a given vertex $m$ equals the number of \textit{arcs} $c(m)$ of this vertex. 
\begin{center}
\includegraphics[scale=0.7]{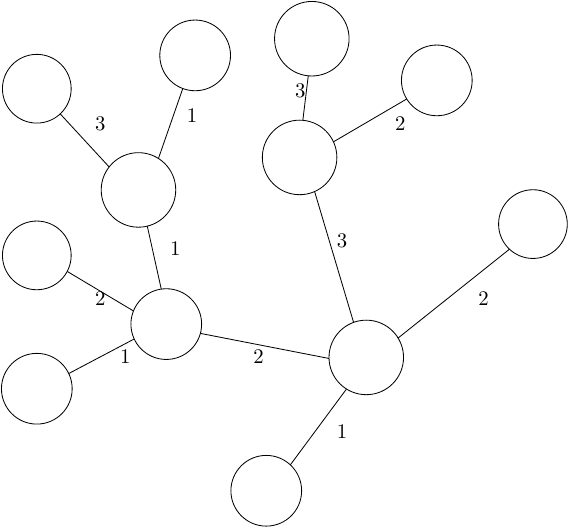} 
\captionof{figure}{A tree contributing to the sum \ref{sumtree}.}
\end{center}\label{fig8}

We close this section with two useful lemmas:
\begin{lemma}\label{lemma1}
For $c(\mathcal{V})>1$, and with the parametrization $\lambda=\rho e^{i\phi}$, $\phi\in [0,\pi/2[$, we have the bound:
\begin{equation}
\bigg|\sum_{\vec{p}\in\mathcal{P}}\bigg(\frac{C_0(\vec{p})}{1-i\sqrt{2\lambda}C_0(\vec{p})\Gamma(\vec{p})}\bigg)^{c(m)}\bigg|\leq K_1(m)\bigg|\frac{1}{\cos(\phi/2)}\bigg|^{c(m)}
\end{equation}
where $K_1(m)$ is a positive constant, depending of the mass parameter in a way that we will precise in the proof.
\end{lemma}
\textbf{Proof}. Observe that we have not precised the domain of $\lambda$. Choosing the parametrization $\lambda=\rho e^{i\phi}$, $\phi\in [0,\pi/2[$, and because $C_0(\vec{p})\Gamma(\vec{p})$ is real:
\begin{equation}
\bigg|\frac{1}{1-i\sqrt{2\lambda}C_0(\vec{p})\Gamma(\vec{p})}\bigg|\leq \bigg|\frac{1}{\cos(\phi/2)}\bigg|,
\end{equation}
implying:
\begin{align}
\bigg|\sum_{\vec{p}\in\mathcal{P}}\bigg(\frac{C_0(\vec{p})}{1-i\sqrt{2\lambda}C_0(\vec{p})\Gamma(\vec{p})}\bigg)^{c(m)}\bigg|&\leq \bigg|\frac{1}{\cos(\phi/2)}\bigg|^{c(m)}\sum_{\vec{p}\in\mathcal{P}}\bigg|\frac{1}{\vec{p}^2+m^2}\bigg|^{c(m)}\\\nonumber
&\leq\bigg|\frac{1}{\cos(\phi/2)}\bigg|^{c(m)}\sum_{\vec{p}\in\mathcal{P}}\bigg|\frac{1}{\vec{p}^2+m^2}\bigg|^{2}.
\end{align}
Since the last sum converges, the lemma is proved.
\begin{flushright}
$\square$
\end{flushright}
\begin{lemma}\label{lemma2}
The sequence $\mathcal{W}_{p}$ verify the bound:
\begin{equation}
|\mathcal{W}_{p}|\leq \bigg|\frac{1}{\cos(\phi/2)}\bigg|K_2(m)
\end{equation}
where $K_2(m)$ is a positive constant.
\end{lemma}
\textbf{Proof}. 
From the definition
\begin{equation}
|\mathcal{W}_{p}|=\sum_{\vec{p}\in\mathcal{P}}\bigg|\frac{C_0(\vec{p})}{1-i\sqrt{2\lambda}C_0(\vec{p})\Gamma(\vec{p})}\delta_{pp_1}\bigg|\leq \bigg|\frac{1}{\cos(\phi/2)}\bigg|\sum_{\vec{p}\in\mathcal{P}}\frac{\delta_{0p_1}}{\vec{p}^2+m^2},
\end{equation}
one more time, because the last sum converges, the lemma is proved\footnote{Note that the sums involving in the proofs of Lemmas \ref{lemma1} and 
 \ref{lemma2} can be computed exactly. For example, using standard complex analysis, one can show that :
\begin{equation*}
\sum_{\vec{p}\in\mathcal{P}}\frac{\delta_{pp_1}}{\vec{p}^2+m^2}=\frac{\pi/2}{\sqrt{3p^2+2m^2}}\coth\big[\pi\sqrt{3p^2+2m^2}\big]\leq \dfrac{\pi/2}{\sqrt{2}m}\coth \big(\sqrt{2}\pi m\big).
\end{equation*}
}. 
\begin{flushright}
$\square$
\end{flushright}
\subsubsection{Convergence}
The presence of the\label{sectionconv} Kronecker deltas $\delta_{p_{t(l)i}p_{s(l)i}}$ in \ref{sumtree} give to the sums the form of a multi-product. More precisely, each loop vertex with $k$ external wavy lines can be represented as a multi-indexd operator $\mathcal{W}_{p_1,...,p_k}$, containing an internal sum over non-external momenta. Then, formula \ref{sumtree} can be written in the form:
\begin{align}\label{structure}
F-F_0=\sum_{\mathcal{T}}\frac{(-2\lambda)^{V(\mathcal{T})-1}}{V(\mathcal{T})!}&\mathcal{N}(\mathcal{T})\int_{0}^1\bigg(\prod_{l\in\mathcal{T}}d\mathit{w}_l\bigg)\Bigg[e^{\frac{1}{2}\sum_{m,p=1}^nX_{mp}(\mathit{w}_l)\frac{\partial}{\tau_{i,m}}\frac{\partial}{\tau_{i,p}}}\\\nonumber
&\qquad\times\prod_{m=1}^{V(\mathcal{T})}(c(m)-1)!\sum_{\{p_{mi}\}}\prod_{m=1}^{V(\mathcal{T})}\mathcal{W}^{(m)}_{p_{m1},...,p_{mk(m)}}\Bigg]_{\tau_{im}=0},
\end{align}
where $\mathcal{N}(\mathcal{T})$ is the number of trees with the same structure but different colors for their intermediate field lines. We distinguish the \textit{leaves} of the tree from the rest. A leaf is a terminal loop vertex involving only one resolvent, and with only one external wavy line. It can be represented by a function of a single variable $\mathcal{W}_{p}$, so that we can write the last term of the previous equation \ref{structure} as:
\begin{equation}
\sum_{\{p_{mi}\}}\prod_{m=1}^{V(\mathcal{T})}\mathcal{W}^{(m)}_{p_{m1},...,p_{mk(m)}}=\sum_{\vec{P}\in\mathbb{Z}^{\mathrm{l}(\mathcal{T})}}\mathcal{A}_{\vec{P}}(\mathcal{T})\prod_{a=1}^{\mathit{l}(\mathcal{T})}\mathcal{W}_{P_a},
\end{equation}
where $\mathit{l}(\mathcal{T})$ is the number of leaves in $\mathcal{T}$, and the big vector $\vec{P}$ lives in the space of external momenta of the leaves : $\mathbb{Z}^{\mathit{l}(\mathcal{T})}$.  $\mathcal{A}_{\vec{P}}(\mathcal{T})$ is the rest of the amplitude, for the tree without its leaves. Thanks to Lemma \ref{lemma2}, the previous sum obeys the bound:
\begin{equation}\label{b1}
\bigg|\sum_{\{p_{mi}\}}\prod_{m=1}^{V(\mathcal{T})}\mathcal{W}^{(m)}_{p_{m1},...,p_{mk(m)}}\bigg|\leq K_2^{\mathit{l}(\mathcal{T})}\bigg|\frac{1}{\cos(\phi/2)}\bigg|^{\mathit{l}(\mathcal{T})}\times \bigg|\sum_{\vec{P}\in\mathbb{Z}^{\mathrm{l}(\mathcal{T})}}\mathcal{A}_{\vec{P}}(\mathcal{T})\bigg|.
\end{equation}
The remaining sum is :
\begin{align}
\bigg|\sum_{\vec{P}\in\mathbb{Z}^{\mathrm{l}(\mathcal{T})}}\mathcal{A}_{\vec{P}}(\mathcal{T})\bigg|&=\bigg|\sum_{\{p_{mi}\}}\prod_{m=1}^{V(\mathcal{T})-\mathit{l}(\mathcal{T})}\mathcal{W}^{(m)}_{p_{m1},...,p_{mk(m)}}\bigg|\leq\bigg|\prod_{m=1}^{V(\mathcal{T})-\mathit{l}(\mathcal{T})}\sum_{\{p_{m1},...,p_{mk(m)}\}}\mathcal{W}^{(m)}_{p_{m1},...,p_{mk(m)}}\bigg|\\
&=\bigg|\prod_{m=1}^{V(\mathcal{T})-\mathit{l}(\mathcal{T})}\sum_{\vec{p}_m\in\mathcal{P}}R(\vec{p}_m)^{c(m)}\bigg|
\end{align}
and thanks to Lemma \ref{lemma1},
\begin{equation}\label{b2}
\bigg|\sum_{\vec{P}\in\mathbb{Z}^{\mathrm{l}(\mathcal{T})}}\mathcal{A}_{\vec{P}}(\mathcal{T})\bigg|\leq K_1(m)^{V(\mathcal{T})-\mathit{l}(\mathcal{T})}\bigg|\frac{1}{\cos(\phi/2)}\bigg|^{\sum_{\{V\in\bar{\mathcal{T}}\}}c(m)}.
\end{equation}
$\sum_{\{V\in\bar{\mathcal{T}}\}}c(m)$ is the number of arcs in the tree $\bar{\mathcal{T}}$, amputated of its leaves. Note that the number of arcs is equal to the number of half wavy lines, or two times the number of wavy lines. Since the number of lines of a tree with $V$ vertices is  $V-1$, we finally deduce that:
\begin{equation}
\sum_{\{V\in\bar{\mathcal{T}}\}}c(m)=2V(\mathcal{T})-2-\mathit{l}(\mathcal{T}).
\end{equation}
\noindent
Grouping together the results \ref{b1} and \ref{b2}, and defining $\mathcal{N}^{\prime}(\mathcal{T})=\mathcal{N}(\mathcal{T})\times\prod_{m=1}^{V(\mathcal{T})}(c(m)-1)!$, we obtain the following bound:
\begin{align}\label{firstbound}
|F-F_0|&\leq \bigg|\sum_{\mathcal{T}}\frac{(-2\lambda)^{V(\mathcal{T})-1}}{V(\mathcal{T})!}\mathcal{N}^{\prime}(\mathcal{T})\big[\sup\big(K_1,K_2\big)\big]^{V(\mathcal{T})}\times \bigg|\frac{1}{\cos^2(\phi/2)}\bigg|^{V(\mathcal{T})-1}\bigg|\\\nonumber
&=\bigg|\frac{2\lambda}{\cos(\phi/2)}\bigg|^{-1}\bigg|\sum_{n=2}^{\infty}\Omega(n)\frac{(2\lambda)^{n}}{n!}\big[\sup\big(K_1,K_2\big)\big]^{n}\times \frac{1}{\cos^{2n}(\phi/2)}\bigg|
\end{align}
where $\Omega(n)$ is a number  depending only on $n$, and defined as:
\begin{equation}
\Omega(n)=3^{n-1}  \sum\limits_{\substack{\{c(m)\}\\\sum_{m=1}^nc(m)=2n-2}}\Omega(n,\{c(m)\})\prod_{m=1}^{n}(c(m)-1)!
\end{equation}
where $\Omega(n,\{c(m)\})$ counts the number of trees with $n$ labeled vertices and coordination numbers $\{c(m)\}$,
and the factor $3^{n-1}$ corresponds to the three possible choices for the color of each intermediate field line, 
From \textit{Cayley's theorem}, we have
\begin{equation}
\Omega(n,\{c(m)\})=\frac{n!}{\prod_{m=1}^{n}(c(m)-1)!}
\end{equation}
hence
\begin{equation}
\Omega(n)=3^{n-1}n!\sum\limits_{\substack{\{c(m)\}\\\sum_{m=1}^nc(m)=2n-2}}1.
\end{equation}
The remaining constrained sum can be easily bounded by the area of the $n-1$ sphere with radius $\sqrt{2n-2}$, which, by Stirling's formula, obeys the bound:
\begin{equation}
\frac{2\pi^{n/2}}{\big(\frac{n}{2}-1\big)!}(2n-2)^{\frac{n-1}{2}}\leq 2\sqrt{2e}\bigg(2\sqrt{\frac{\pi}{e}}\bigg)^{n-1},
\end{equation}
so that:
\begin{equation}
\Omega(n)\leq 2\sqrt{2e}\,3^{n-1}\bigg(2\sqrt{\frac{\pi}{e}}\bigg)^{n-1}n!
\end{equation}
and \ref{firstbound} becomes
\begin{align}\label{finalbound}
\nonumber|F-F_0|&\leq 2\sqrt{2e}\bigg|\frac{12\sqrt{\pi/e}\lambda}{\cos^2(\phi/2)}\bigg|^{-1}\bigg|\sum_{n=2}^{\infty}(12\sqrt{\pi/e}\lambda)^{n}\big[\sup\big(K_1,K_2\big)\big]^{n}\times\frac{1}{\cos^{2n}(\phi/2)}\bigg|\\
&\leq\bigg(\frac{6|\lambda|}{\cos^2(\phi/2)}\bigg)^{-1}\sum_{n=2}^{\infty}(|\lambda|K)^{n},
\end{align}
with:
\begin{equation}\label{K}
K:=\frac{12\sqrt{\pi/e}\sup\big(K_1,K_2\big)}{\cos^2(\phi/2)}.
\end{equation}
Finally, the first term $F_0$ is trivially bounded. Indeed, with integral representation of logarithm and a partial integration over intermediate fields, we find:
\begin{align}
\bigg|\int d\nu_{X(\mathit{w}_l)}(\tau)\sum_{\vec{p}\in\mathcal{P}}\ln(1-i\sqrt{2\lambda}&C_0(\vec{p})\Gamma(\vec{p}))\bigg|\\
&=\bigg|\int d\nu_{X(\mathit{w}_l)}(\tau)\int_{0}^1dt\sum_{\vec{p}\in\mathcal{P}}\frac{-i\sqrt{2\lambda}C_0(\vec{p})\Gamma(\vec{p})}{1-i\sqrt{2\lambda}tC_0(\vec{p})\Gamma(\vec{p})}\bigg|\\\nonumber
&=3\bigg|\int d\nu_{X(\mathit{w}_l)}(\tau)\int_{0}^1dt\sum_{\vec{p}\in\mathcal{P}}\frac{2\lambda [C_0(\vec{p})]^2t}{(1-i\sqrt{2\lambda}tC_0(\vec{p})\Gamma(\vec{p}))^2}\bigg|,
\end{align}
so that using Lemma \ref{lemma1}, convergence is easy. Hence, we have proved the first result of this paper:
\begin{theorem}
The free energy expansion is absolutely convergent for a small enough coupling. More precisely, the analyticity domain is 
defined by the equation
$|\lambda|<\frac{1}{K}$, which, with definition \ref{K} for $K$, corresponds to the interior of a cardioid in the complex plane. 
\end{theorem}

\subsubsection{Borel summability}

After the convergence, we now move on to the Borel summability of the perturbative expansion. The previous result shows that the first requirement of Theorem \ref{Borel} is satisfied, because we can find a disk inside of the cardioid in which the series converge, and we will establish the second point in this section. Calling $
R_rF(\lambda)$ the Taylor remainder of order $r$ for the free energy $F$:
\begin{equation}\label{rest}
R_rF(\lambda):=\lambda^{r+1}\int_{0}^1\frac{(1-t)^r}{r!}F^{(r+1)}(t\lambda)dt.
\end{equation}
Expanding with the tree formula, we get:
\begin{equation}\label{restF}
R_rF(\lambda):=\sum_{n=1}^{\infty}\frac{2^{n-1}}{n!}\sum_{\mathcal{T}_n}\int_{0}^1\bigg(\prod_{l\in\mathcal{T}_n}d\mathit{w}_l\bigg)\int d\nu_{X(\mathit{w}_l)}(\tau)R_r[\mathcal{Y}_{\mathcal{T}_n}],
\end{equation}
where:
\begin{equation}\label{Y}
\mathcal{Y}_{\mathcal{T}_n}:=(-\lambda)^{n-1}\prod_{m=1}^{n}\sum_{\vec{p}_m\in\mathcal{P}}
\prod_{m=1}^n(c(m)-1)!R(\vec{p}_m)^{c(m)}\prod_{l\in\mathcal{T}_n}\bigg(\sum_{i=1}^3\delta_{p_{t(l)i}p_{s(l)i}}\bigg).
\end{equation}
When $n-2\geq r$, $R_r[\mathcal{Y}_{\mathcal{T}_n}]=\mathcal{Y}_{\mathcal{T}_n}$. In this case, each term of the sum has a bound of the form $K^{n}|\lambda|^n$, and the sum converges absolutely for small enough coupling. For $n-2<r$, however, the remainder is obtained by a Taylor expansion of the resolvents involved in $\mathcal{Y}_{\mathcal{T}}$. We can extract the factor $\lambda^{n-1}$ in front of $\mathcal{Y}_{\mathcal{T}}$, and write: $\mathcal{Y}_{\mathcal{T}_n}=\lambda^{n-1}\bar{\mathcal{Y}}_{\mathcal{T}_n}$, so that: $R_r[\mathcal{Y}_{\mathcal{T}_n}]=\lambda^{n-1}R_{r-n+1}[\bar{\mathcal{Y}}_{\mathcal{T}_n}]$. Introducing $z = i \sqrt{2 \lambda}$, since
\begin{equation}
r_z(\vec{p})=\frac{1}{1-z(C_0\Gamma)(\vec{p})},
\end{equation}
we have
\begin{equation}
\frac{\partial^n}{\partial z^n}r_z(\vec{p})=(C_0\Gamma)^n(\vec{p})n!r_z^{n+1}(\vec{p}),
\end{equation}
and the Taylor expansion of  formula \ref{Y} in powers of $z$ leads to:
\begin{align}
\bar{\mathcal{Y}}_{\mathcal{T}_n}=\sum_{k=0}^{\infty} &\frac{z^k}{k!} \prod_{m=1}^n \sum_{\{k_l\}|\sum_{k_l}=k}\frac{k!}{\prod_{l=1}^n k_l!}\\
&\times\sum_{\vec{p}_m}\big[(C_0\Gamma)^{k_m}C_0^{c(m)}\big](\vec{p}_m)(c(m)+k_m)!\prod_{l\in\mathcal{T}_n}\bigg(\sum_{i=1}^3\delta_{p_{t(l)i}p_{s(l)i}}\bigg)
\end{align}
and from formula \ref{rest}, we find:
\begin{align}\label{sumtree2}
R_{r-n+1}[\bar{\mathcal{Y}}_{\mathcal{T}_n}]= & z^{2(r-n)}\int_{0}^1dt\frac{(1-t)^{2r-2n+2}}{(2r-2n+2)!}  \sum_{\{k_l\}|\sum_{k_l}=2r-2n+3}\frac{(2r-2n+3)!}{\prod_{l=1}^n k_l!}\prod_{m=1}^n\\\nonumber
&\times\sum_{\vec{p}_m}\big[(C_0\Gamma)^{k_m}C_0^{c(m)}\big](\vec{p}_m)r_{tz}^{c(m)+k_m}(\vec{p}_m)(c(m)+k_m)!\prod_{l\in\mathcal{T}_n}\bigg(\sum_{i=1}^3\delta_{p_{t(l)i}p_{s(l)i}}\bigg).
\end{align}
We can now report this expression in equation \ref{restF}. First, note that because $C_n^p\leq 2^n$, the factor $c(m)(c(m)+k_m)!/[k_m!c(m)!]$ is bonded by $e^{\ln(c(m))}2^{c(m)+k_m}\leq (2e)^{c(m)}2^{k_m}$, and the product over $m$ gives a factor $(2e)^{2n-2}2^{2r-2n+3}$. Secondly, we can perform the Gaussian integration. Because there are $2r-2n$ fields, the number of Wick contractions is $(2(r-n)!! = 2^{r-n}(r-n)!\leq 2^rr!$. These contractions add many intermediate loop lines to the original tree. 
A typical contribution to the Wick contractions can be pictured as in figure \ref{fig9} below. 
\begin{center}
\includegraphics[scale=0.7]{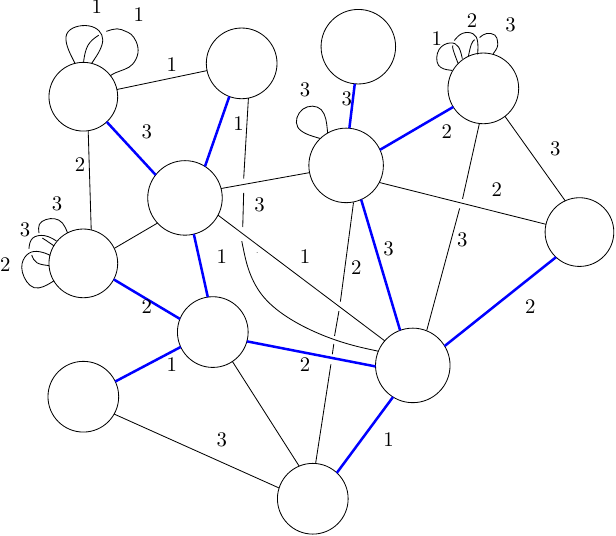} 
\captionof{figure}{Typical contribution of the sum \ref{sumtree2}, with loops and tadpoles. The \textit{root tree} lines are pictured in blue.}
\end{center}\label{fig9}
In this picture, which is based on the tree of figure \ref{fig8}, that we call \textit{root tree}, the  original tree is enriched with additional loop lines coming from the Wick contraction of the intermediate fields which were hidden in the tree resolvents. Each vertex has coordination $c(m)+k_m$, where  $c(m)$ is the original number of arcs. The $k_m$ intermediate fields of a given vertex $m$ can be contracted together, giving tadpoles, or with the intermediate fields of others vertices, forming the loops. From Lemma \ref{lemma1}, it follows that $|r_{tz}|\leq |\cos(\phi/2)|^{-1}$. Moreover, the contribution of the intermediate fields is twofold. The $C^{k_m}(\vec{p}_m)$ decrease the weight of each vertex, and the additional lines can be viewed as constraint on the sums over $\vec{p}_m$. As a result, all the contributions are bounded by one of the root tree, giving the bound:
\begin{equation}
(2(r-n)!!\big[\sup(K_1,K_2)\big]^{n}\bigg|\frac{1}{\cos(\phi/2)}\bigg|^{4n-2+(2r-2n)}\leq 2^rr!\bigg[\frac{\sup(K_1,K_2)}{\cos^2(\phi/2)}\bigg]^n\bigg|\frac{1}{\cos^2(\phi/2)}\bigg|^{r-1}.
\end{equation}
Finally, the remaining integration over $t$ gives $\int_{0}^1(1-t)^{2r-2n+2}=1/(2r-2n+3)$, which together with the denominator $(2r-2n+2)!$ exactly compensates the combinatorial factor $(2r-2n+3)!$. As in the previous section, using Cayley's theorem for the number of trees with $n$ vertices and Stirling's formula, as in \ref{sectionconv}, we find a bound of the form: $AB_1^nB_2^rr!$ for some constants $A$, $B_1$ and $B_2$. Because $n-2<r$, summing over $n$, we find the final bound : $A'|\lambda|^rB^rr!$ for the contributions in \ref{restF} for which $n-2<r$. As explained before, the contributions for $n-2\geq r$ are all bounded by bounds of the form : $|\lambda|^{n}K^{n}$, and the sum behaves as : $A''|\lambda|^rK^r$. Ultimately, because, for positive constants $k_1$ and $k_2$, $k_1r!+k_2\leq (k_1+k_2)r!$, we find that $|R_rF(\lambda)|\leq A''(B')^r|\lambda|^rr!$, which corresponds to the second condition of Theorem \ref{Borel}. It completes the proof of Borel summability.

\subsection{Schwinger functions}

Schwinger functions, or connected correlation functions are obtained from the logarithm of \ref{intermediatevec} by functional derivative with respect to the sources $J$ and $\bar{J}$. More precisely, the connected function $S_{2N}(\{\vec{p}_i,\vec{\bar{p}}_i\})$ with $2N$ external lines and external momenta $\vec{p}_i$ and $\vec{\bar{p}}_i$, $i=1,...,N$, is given by:
\begin{equation}\label{connectedSF}
S_{2N}(\{\vec{p}_i,\bar{\vec{p}}_i\}):=\prod_{i=1}^N\frac{\partial}{\partial J(\vec{\bar{p}}_i)}\frac{\partial}{\partial \bar{J}(\vec{p}_i)}\ln\big(\mathcal{Z}[J,\bar{J}]\big)\big|_{J,\bar{J}=0}.
\end{equation}
Expanding $\ln\big(\mathcal{Z}[J,\bar{J}]\big)$ with the help of the forest formula, as in the previous section for the free energy, we find that the term that we have called $W$ includes the source term involved in \ref{intermediatevec}.  Because of \ref{connectedSF}, only the terms with $N$ such source terms give a non-zero contribution. This contribution is a resolvent. The remaining terms, involving the logarithm are then derived, but the connectivity impose that the resolvent factors coming from the derivative of the source terms are derived at least one time with respect to the $\tau_i$. Once more, using Lemma \ref{lemma1} each resolvent or derivative of resolvent can be bounded by a constant in the cardioid domain, in the same way as for the free energy, and the absolute convergence of the expansion of any Schwinger function follows. \\

\section{Conclusion}

We have shown that an ultraviolet convergent TGFT can be successfully treated with the LVE, and proved the existence of a finite analyticity domain in which perturbative expansion makes sense. As we saw, the closure constraint considerably simplified the proofs in comparison to similar works for models without this additional symmetry \cite{constructivetensor2}. Hence, if the result obtained in this paper is not very surprising, it is an encouragement for future works on more complicated TGFTs with divergences and renormalization. The next step in the constructive program for TGFTs should be to construct \textit{super-renormalizable} theories, such as $U(1)$-$T^4_4$ in dimension $4$ and $U(1)$-$T^4_5$ 
in dimension 5. Such theories need only the renormalization of a finite set of divergent graphs. However, even in this simple case, taking properly into account the corresponding subtractions in a constructive way requires a refined technique, the so-called \textit{Multi-scale Loop Vertex Expansion} (MLVE) which includes a multi-scale decomposition. 
A further more difficult step should be the construction of  a \textit{just-renormalizable} TGFT such as $U(1)$-$T^4_6$, which is asymptotically free \cite{Lahoche:2015ola}. 

\section{Acknowledgments}
The author specially thanks his supervisor Vincent Rivasseau for his suggestions at the initiative of this project, his advice and his patient proofreading; and his wife Julie Lahoche for the languages corrections.

\end{document}